\documentclass[apjl]{emulateapj}
\slugcomment{Accepted for publication in Astrophysical Journal Letters}
\newcommand\asca{{\it ASCA}}

\newcommand\xmm{{\it XMM-Newton }}

\newcommand\kev{{\rm~keV}}

\newcommand\kms{\ifmmode {\rm~km\ s}^{-1} \else ~km s$^{-1}$\fi}
\newcommand\Hunit{\ifmmode {\rm~km\ s}^{-1}\ {\rm Mpc}^{-1}
        \else ~km s$^{-1}$ Mpc$^{-1}$\fi}
\newcommand\ctssec{\ifmmode {\rm~count\ s}^{-1} \else ~count s$^{-1}$\fi}
\newcommand\ergsec{\ifmmode {\rm~erg\ s}^{-1} \else
        ~erg s$^{-1}$\fi}
\newcommand\funit{\ifmmode {\rm~erg\ s}^{-1}\;{\rm cm}^{-2} \else
        ~ergs s$^{-1}$ cm$^{-2}$\fi}
\newcommand\phflux{\ifmmode {\rm~photon\ s}^{-1}\;{\rm cm}^{-2}
        \else   ~photon s$^{-1}$ cm$^{-2}$\fi}
\newcommand\efluxA{\ifmmode {\rm~erg\ s}^{-1}\;{\rm cm}^{-2}\;{\rm
        \AA}^{-1} \else ~erg s$^{-1}$ cm$^{-2}$ \AA$^{-1}$\fi}
\newcommand\efluxHz{\ifmmode {\rm~erg\ s}^{-1}\;{\rm cm}^{-2}\;{\rm
        Hz}^{-1} \else ~erg s$^{-1}$ cm$^{-2}$ Hz$^{-1}$\fi}
\newcommand\cc{\ifmmode {\rm~cm}^{-3} \else cm$^{-3}$\fi}
\newcommand\FWHM{\ifmmode {\rm~FWHM} \else ${\rm~FWHM}$\fi}
\newcommand\Msun{\ifmmode M_{\odot} \else $M_{\odot}$\fi}
\newcommand\Lsun{\ifmmode L_{\odot} \else $L_{\odot}$\fi}

\newcommand\hbeta{\ifmmode {\rm H}\beta \else H$\beta$\fi}
\newcommand\Kalpha{\ifmmode {\rm K}\alpha \else K$\alpha$\fi}
\newcommand\nh{\ifmmode N_{\rm H} \else N$_{\rm H}$\fi}

\shorttitle{Soft time lags in Mrk 1040}
\shortauthors{Tripathi et al.}

\begin{document}

\title{Soft time lags in the X-ray emission of Mrk 1040}

\author{Shruti Tripathi\altaffilmark{1},  Ranjeev Misra\altaffilmark{2}, Gulab Dewangan\altaffilmark{2} \& Shantanu Rastogi\altaffilmark{1} }

\altaffiltext{1}{Dept. of Physics, DDU Gorakhpur University, Gorakhpur
273009, India; email: stripathi@iucaa.ernet.in}

\altaffiltext{2}{Inter-University Centre for Astronomy and
Astrophysics, Pune 411007, India; email: rmisra@iucaa.ernet.in}

\begin{abstract}
  Temporal analysis of X-ray binaries and Active Galactic Nuclei have
  shown that hard X-rays react to variation of soft ones after a time
  delay. The opposite trend, or soft lag, has only been seen in a few
  rare Quasi-periodic Oscillations in X-ray binaries and recently for
  the AGN, 1H 0707-495, on short timescales of $\sim 10^3$
  secs. Here, we report analysis of a \xmm{} observation of Mrk~1040,
  which reveals that on the dominant variability timescale of $\sim
  10^4$ secs, the source seems to exhibit soft lags. If the lags are frequency
  independent, they could be due to reverberation effects of a
  relativistically blurred reflection component responding to a
  varying continuum. Alternatively, they could be due to
  Comptonization delays in the case when high energy photons impinge
  back on the soft photon source.  Both models can be verified and
  their parameters tightly constrained, because they will need to
  predict the photon spectrum, the r.m.s variability and time lag as a
  function of energy. A successful application of either model will
  provide unprecedented information on the radiative process, geometry
  and more importantly the size of the system, which in turn may
  provide stringent test of strong general relativistic effects.

\end{abstract}

\section{Introduction}

Active Galactic Nuclei (AGN) are known to show large amplitude variability
in X-rays. The variability broadly correlates with the black hole
mass \citep[e.g.][]{One05}, but perhaps a more unifying result is that
the break in the power spectrum depends on the black hole mass and
luminosity in a similar way for both AGN and Galactic black hole systems,
providing a fundamental plane of these measured quantities \citep{Mch06}.

Study of the variability as a function of energy, especially time lags
between different energy bands, has provided insight into the nature
of the variability, the radiative processes and geometry of these
black hole systems.  For Galactic black hole systems (e.g. Cyg X-1)
the lags are found to be proportional to the logarithm of the energy
bin and more importantly they depend on the frequency of the
variability \citep{Now99}.  Similarly, all AGN with well measured time
lag spectra show that the variations in the hard band lag behind those
in the soft band and the size of the hard lag decreases with Fourier
frequency \citep{Papa01,Vaug03,Mch04,Are06,Mark07,Are08}. The energy
dependent time lags in Mrk~110 were found to be proportional to the
logarithm of the energy \citep{Das06}. These observations demonstrate
the similarity of the nature of X-ray variability in the two different
mass scales - X-ray binaries and AGN.
The frequency dependence of the time lags cannot be reconciled with a
model where they are produced by multiple Compton scattering. Instead,
cylindrical inward propagation of fluctuations in the disk \citep{Lyu97},
can more naturally explain the dependence. The model requires
non-localized origin  of the hard and soft photons which may be achieved
either by a transition disk region \citep{Mis00} or spectral hardening
in the inner regions \citep{Kot01,Are06}.
   
These time lags reported are hard i.e. the hard X-rays react after a delay
to soft ones. For Galactic X-ray binaries (including Neutron stars), the
opposite trend of a soft lag has only been reported for  kHz Quasi-periodic
Oscillations \citep{Vau98,Kaa98} and for alternate harmonics of a low frequency
QPO in the black hole system GRS 1915+105 \citep{Cui99} and seemingly never
for the broad band continuum variability. Recently, \cite{Fab09} report a soft
lag in the AGN 1H 0707-495, between
$0.3$-$1.0$ and $1.0$-$4.0$ keV bands 
on short timescales of $\sim 1000$ secs. However, the dominant variability
on longer timescales exhibits  typical hard lags. Interestingly,
the source spectrum can be modeled with a complex reflection component
(with both Iron K and Iron L emission lines) and the soft lag could be
due to reverberation effects between the continuum and Iron L line
\citep{Fab09}. In general, the temporal analysis so far undertaken
for AGN and Galactic black holes, reveal that for the dominant variability,
the hard photons lag the soft.

\asca{} observation of Mrk~1040 reveals a rather typical variable AGN
\citep{One05,Tur99} with its spectrum showing Iron line emission
and perhaps warm absorber signatures \citep{Ren95}. Here we report
analysis of a \xmm observation and the surprising result is that in the dominant variability timescale of $10^4$ secs, the source seems to
exhibit soft lags.

\section{Observation, Analysis \& Results}

\xmm observed Mrk~1040 on 13 February, 2009 for the
duration of 90.9 ksec (observation ID 0554990101).  
We processed and filtered the EPIC-pn data in a standard way using the 
 Science Analysis Software (SAS) version 10.0 and using the recent calibration files. 
We examined the full field lightcurve above $10\kev$ for flaring particle background which were identified and excluded. We used a continuous exposure of $70.4{\rm~ks}$ for temporal analysis. For spectral analysis we used the same length ofdata as taken in temporal study. We extracted source and background spectra from a source-centered circular region of radius $36\arcsec$ and  an off-source region, respectively, and using single and double events with FLAG=0. We also extracted lightcurves using the above criteria and additionally using quadruple events.  
Response matrices and auxiliary files
were generated using the RMFGEN and ARFGEN tools. The EPIC-pn spectrum
was grouped using the SAS task specgroup. 

\begin{figure}
\begin{center}
{\includegraphics[width=1.0\linewidth,angle=0]{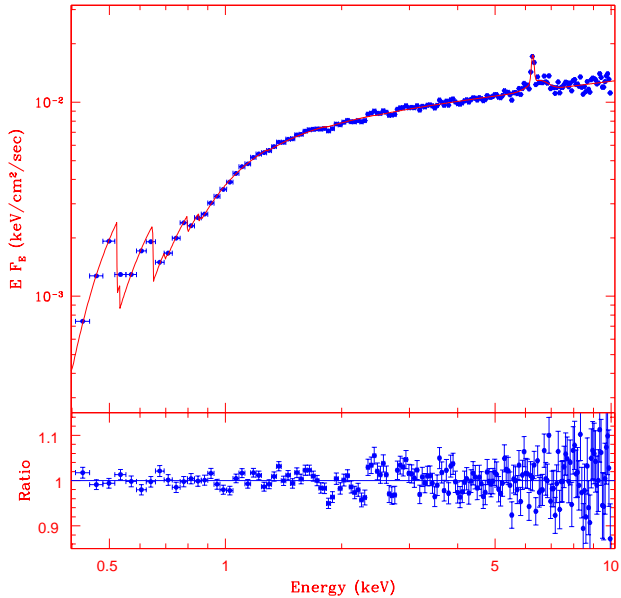}}
\end{center}
\caption{The unfolded spectrum of Mrk 1040 fitted by an empirical model
consisting of absorbed  dual-Comptonization components and some prominent
line and edges (see text for details). The ratio of the data to the model
shows several systematic features but the model is a reasonable
first approximation at a few percentage level to the overall spectrum. }
\label{spec}
\end{figure}

As is the case for many AGN, 
the time-averaged X-ray spectrum  of Mrk 1040 as observed by \xmm is complex.
There is intrinsic neutral absorption, a soft excess feature, multiple 
Iron line emissions, plus edges and absorption features, probably due to
a warm absorber. The complex Iron line features indicate the possibility of
a complex blurred reflection component. 
Moreover, as shown below, the source exhibited significant
secular spectral evolution during the observation and hence steady state
models may not be appropriate to describe the time-averaged spectrum. Our
motivation here is not to fit the spectrum in detail, but to get a first
order approximation which describes its overall shape.

\begin{figure}
\begin{center}
{\includegraphics[width=1.0\linewidth,angle=0]{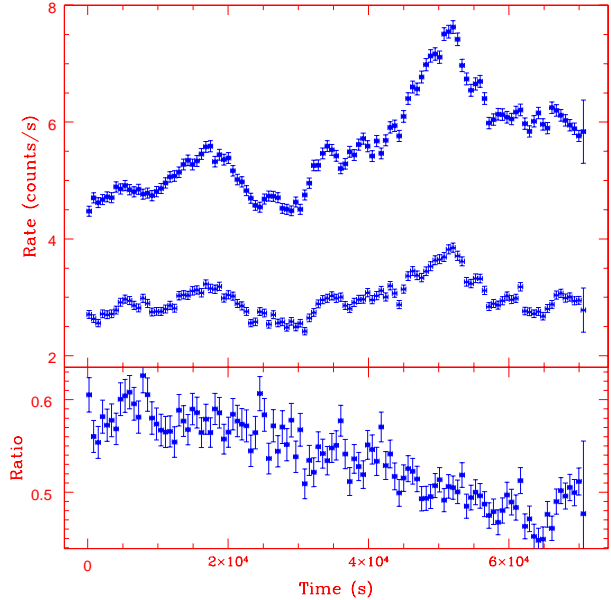}}
\end{center}
\caption{Top panel: The $640$ sec binned X-ray lightcurves of Mrk 1040 
in the soft, 
$0.2$-$2.0$ keV (circles) and hard, $2.0$-$10.0$ keV (triangles) 
energy bands. Bottom panel:
The ratio of the hard to soft count rates.    }
\label{lcurve}
\end{figure} 

\begin{figure}
\begin{center}
{\includegraphics[width=1.0\linewidth,angle=0]{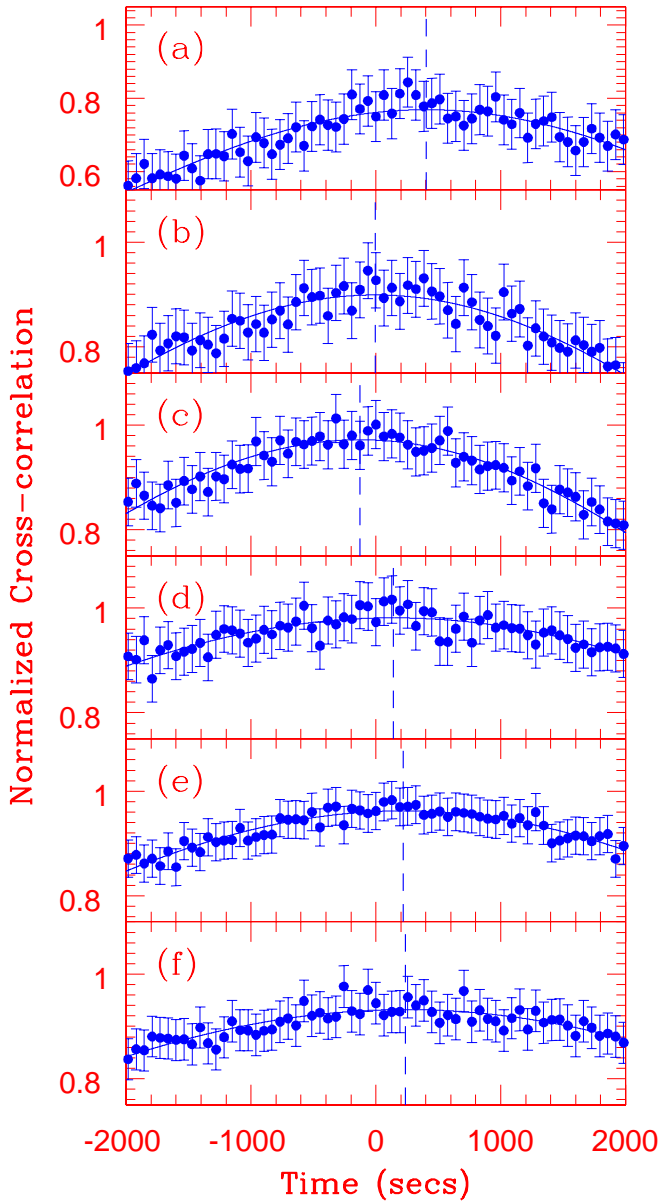}}
\end{center}
\caption{The cross-correlation function for different energy bands. The
reference energy band is $1.0$-$2.0$ keV. From top to bottom: 
(a) $5.0$-$10.0$ (b) $3.0$-$5.0$  (c) $2.0$-$3.0$  (d) $0.8$-$1.0$  
(e) $0.5$-$0.8$ (f) $0.3$-$0.5$. The solid line is the best fit Gaussian
curve and the dashed vertical lines mark the centroid.     }
\label{cross}
\end{figure}

\begin{figure}
\begin{center}
{\includegraphics[width=1.0\linewidth,angle=0]{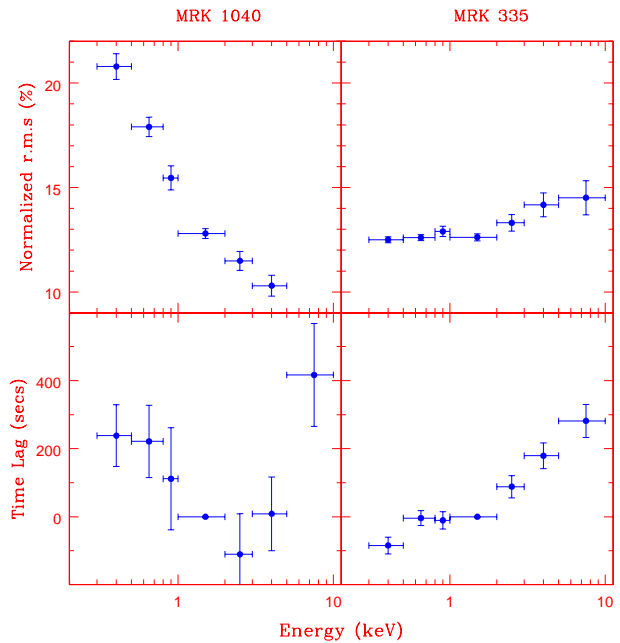}}
\end{center}
\caption{The r.m.s variability and the time lag as a function of energy
for Mrk~1040 (left) and Mrk~335 (right column). An
identical analysis has been undertaken for both the sources. The errorbars in
the time lags are 1-sigma levels estimated by simulations.    }
\label{cross_comp}
\end{figure}

We model the spectrum
using an intrinsically absorbed dual Comptonization model. One Comptonization
component to describe the soft excess and the other for the hard X-ray 
continuum. To this basic model we include some prominent line emission and
edges. The Galactic absorption towards the direction of Mrk 1040 is
$\sim 6.64 \times 10^{20}$ cm$^{-2}$ and the data requires an additional
intrinsic neutral absorption of $\sim 2.0\pm0.2 \times 10^{21}$ cm$^{-2}$.
The soft excess is described by a Comptonized component represented
by the XSPEC model ``nthcomp'' \citep{Zdz96,Zyc99} with spectral 
index $\Gamma_S = 3.1\pm0.2$ and 
electron temperature $kT_e = 0.34\pm0.04$ keV. 
A second Comptonization model for
the hard X-ray emission, is represented by the XSPEC convolution model 
``Simpl'' \citep{Ste09} acting on the first Comptonization 
model ``nthcomp''. Thus, in
this description, the first Comptonization component which gives rise to
the soft excess, is the origin of the seed photons for the second 
Comptonization component giving the hard X-ray continuum. The high energy
spectral index is $\Gamma= 1.73\pm0.02$ and the 
fraction of soft excess photons getting scattered is 
$f = 0.05\pm0.03$. The Iron line emission feature is modeled as a narrow
Gaussian at $6.36\pm0.02$ keV and a broad one centered at $6.6\pm0.1$ keV with
a width of $0.4\pm0.1$ keV. Two prominent edges at $0.66\pm0.005$ and 
$0.81\pm0.01$ keV with optical 
depth $0.72\pm0.05$ and $0.21\pm0.03$ respectively have been included. For
a distance of $51$ Mpc, the observed and unabsorbed luminosities are
$1.2$ and $1.9 \times 10^{43}$ ergs/s respectively. The model is formally
unacceptable with a $\chi^2/d.o.f =281/161$. This is evident in 
Figure \ref{spec} where the unfolded spectrum and the ratio of the
data to model is shown. There are systematic variations in the ratio
plot at several energies which may be due to the presence of complex emission/
absorption or blurred reflection features and/or spectral variability during
the observation. However, as can be seen in Figure \ref{spec} the model is
a good first order approximation of the data at a few percentage level. 
The essence of this spectral analysis is that with a high energy
spectral index of $\sim 1.7$, moderate intrinsic absorption and soft excess,
the spectrum of Mrk 1040 is typical of AGNs of its class.

The top panel of Figure \ref{lcurve} shows the background subtracted 
X-ray $640$ sec binned lightcurves of Mrk~1040 in soft, $0.2$-$2.0$ keV 
(circles) and 
hard, $2.0$-$10.0$ keV (triangles) energy bands. There is clear
strong correlated variability in the two bands on  $\sim 10^4$ sec timescales.
The bottom panel shows 
that the ratio of the hard to soft count rates (i.e. the hardness ratio) 
significantly decreases
from $\sim 0.6$ to $\sim 0.45$ during the course of the observation,
indicating spectral evolution of the source on long ($\sim 10^5$ sec) 
timescales. It is interesting to note that the variation in the
hardness ratio is clearly uncorrelated with the variability of the
lightcurves. This means that the primary variability of the source
in $\sim 10^4$ sec timescales is not associated with significant spectral
variation and hence its origin cannot be due to changes in the absorption 
medium which would effect primarily the soft band.

Lightcurves binned at $64$ secs were generated for seven energy bands
corresponding to $0.3$-$0.5$, $0.5$-$0.8$, $0.8$-$1.0$, $1.0$-$2.0$, 
$2.0$-$3.0$, $3.0$-$5.0$ and $5.0$-$10.0$ keV. Cross-correlation analysis
was undertaken between the reference band $1.0$-$2.0$ keV and others. 
The XRONOS task ``crosscor'' was used for the analysis. The functions
were normalized by dividing by the square root of the product of 
the excess variances of the two series (i.e. with  ``crosscor'' flag 
``norm = 2''). The entire lightcurve was taken as a single segment and
the analysis was undertaken in the ``slow'' mode (i.e. not in fast Fourier
transform mode). In such a case, the task computes errors on the
cross-correlation by propagating the measurement errors of each time bin.
Dividing the light-curve into five or more segments, gave negative
variance for some segments, rendering the analysis unfeasible.  
The cross-correlations in the delay time range of $-2000$ to $2000$ secs
are plotted in Figure \ref{cross}. The lightcurves are well
correlated with the cross-correlation function peaking at values $> 0.8$. 
In this delay range,
we fit the functions with a Gaussian and obtain the best fit centroid. 
The best fit Gaussians are plotted in the figure (solid lines)
and their centroid values are marked with a horizontal dashed line which
is taken as the time-delay between the lightcurves.  

We estimate the significance of the measured time-delays  through simulations.
The intrinsic power spectrum of the source is an approximate power-law with index
$\sim 2$. Based on this power spectral shape and using the method proposed by
\cite{Tim95}, we simulate 200 pairs of lightcurves having the same measured time-delay as the observed pair.
Measurement errors were added to each lightcurve and the
pairs were then subjected to an identical analysis, where the cross-correlation function was
fitted by a Gaussian function.  The root mean square deviation of
the centroids of the best fit Gaussian functions, was then taken to be the 1-sigma error on the time-delay.

The 
first column of Figure \ref{cross_comp} shows the r.m.s and the
time lag for Mrk 1040 are plotted as function of energy bin.
Below $2$ keV, the time lag decreases with energy which means that
the variation in the soft photons occur {\it after} the corresponding variation
in the hard ones i.e. there is a soft time lag.  For energies $> 2$ keV, the time lag increases with energy as
in the regular case of hard lags. To validate the analysis technique and to make a direct comparison with
other sources, the  right column of Figure \ref{cross_comp} show
the r.m.s and time lag for Mrk 335, computed using an
{\it identical} analysis of its similar length \xmm observation.
Although the r.m.s varies differently with energy, the time lag for
Mrk~335 increases with energy (i.e. hard lags). This is
consistent with the more detailed analysis, including frequency dependence,
of this source \citep{Are08}.

\section{Discussion}

Figure   \ref{cross_comp} shows that the r.m.s 
decreases with energy for Mrk 1040, while it 
is nearly constant for Mrk~335. This may indicate that
the variability in Mrk 1040 is due to variations in the absorbing medium. 
Moreover,
if the absorbing medium (e.g. warm absorber) reacts to a change in the
hard X-ray continuum with a time delay, this could naturally explain the
soft lags observed in the source. However, as seen in Figure \ref{lcurve},
the hardness ratio variation is on longer timescale and uncorrelated
with the intensity variation. Thus, such a model cannot explain the
soft lags observed.

Frequency dependent time lags in X-ray binaries and AGN can be 
explained in terms of fluctuations
propagating from the outer regions of the disk to the inner 
\citep{Lyu97}.  To reconcile soft lags,
either the waves have to propagate outwards or the hard photons
have to arise from outer regions, both of which seem rather physically
unrealistic.

In the most straight forward interpretation,  Comptonization  naturally
predicts a time lag between energy bands to be $\sim R/c$ $\log(E_2/E_1)$, where
$R$ is the size of the region and $E_2/E_1$ is the ratio of the energies. While
the energy dependence is consistent with what is observed in black hole
binaries and AGN, the Comptonization interpretation is often ruled out because
it is difficult to reconcile with the observed frequency dependence of the
lags. However, it is important to note that Comptonization lag must 
exist and should manifest at high enough frequencies when the wave propagation
lag is small. Thus at high frequencies the lag should saturate to the
Comptonization lag values. While, this saturation
 has not been detected, the observed time lags already impose a
stringent upper limit on the size of the Comptonizing region, $R$ for AGNs.
For Ark 564, the time lag variation of $\sim 50$ sec $\log (E_2/E_1)$ 
\citep{Are06} requires that $R < 2 \times 10^{11}$ cm or $ < 2 GM/c^2$ for
at $10^7 M_\odot$  black hole.   
  
It is not known whether the lags in Mrk 1040 are frequency dependent and
hence a Comptonization origin may still be viable. Soft lags due to
Comptonization is indeed possible, as invoked  to explain the soft lags
observed for kHz QPO in X-ray binaries \citep{Lee01}. A fluctuation in the
electron temperature will lead to variations in the hard X-rays after a
delay. A fraction of these hard X-rays may impinge back on the input photon producing region and hence affect the soft photons. This will lead to a detectable soft lag. 
Alternatively, the soft lag could be due to reverberation of a complex
gravitationally blurred reflection component to variations of the continuum,
as proposed for 1H 0707-495 \citep{Fab09}. It is interesting to note that
if like for 1H 0707-49, the continuum in Mrk~1040 only dominates 
in the $1$-$2$ keV band, the time lag variation with energy observed, may be
naturally explained. The temporal and spectral signatures of both these models
can be quantified, although, especially for the reflection scenario, the
complex and non-intuitive effect of light bending needs to be taken into
account. Both models can be tested and their parameters tightly constrained,
because they will need to self-consistently explain the lag and r.m.s versus
energy as well as the photon spectrum. Successful application of either
model will provide rich dividends in terms of constraining the radiative
processes, geometry and more importantly the size of the system and provide
opportunity to test strong General Relativistic effects. Note however, that
for both scenarios, the time lag should not be frequency dependent.

\end{document}